\documentstyle[aps,prd,t1enc,amsmath,epsfig]{revtex}
\begin{document}
\draft

% Two-column format, abstract excluded.
\twocolumn[ 
\hsize\textwidth\columnwidth\hsize\csname@twocolumnfalse\endcsname 

\title{Photon frequency conversion induced by gravitational radiation}

\author{Gert Brodin,$^1$
  Mattias Marklund$^{2,3}$ and
  Martin Servin$^1$}  

\address{$1$ Department of Plasma Physics,
  Ume{\aa} University, SE--901 87 Ume{\aa}, Sweden}
\address{$2$ National Defence Research
  Establishment FOA, SE--172 90 Stockholm, Sweden}
\address{$3$ Department of Electromagnetics,
  Chalmers University of Technology, SE--412 96 G\"oteborg, Sweden}

\date{\today}

\maketitle

\begin{abstract} 
We consider propagation of gravitational radiation in a
magnetized multicomponent plasma. It is shown that large density
perturbations can be generated, even for small deviations from flat
space, provided the cyclotron frequency is much larger than the plasma
frequency. Furthermore, the induced density gradients can generate
frequency conversion of electromagnetic radiation, which may give rise
to indirect observational effect of the gravitational waves.
\end{abstract}

\pacs{04.30.Nk, 52.35.Mw, 95.30.Sf} 

] % Ending of two-column format

%%%%%%%%%%%%%%%%%%%%%%%%%%%%%%% 

\section{Introduction}

%%%%%%%%%%%%%%%%%%%%%%%%%%%%%%% 

Recently there has been an increased interest in gravitational waves, mainly
due to possibility of direct detection by LIGO (Laser Interferometer
Gravitational-Wave Observatory) \cite{LIGO}. Naturally the effects of
gravitational waves on earth are very small -- which is illustrated by the
large dimensions required for detection. Closer to the source the influence
of the gravitational waves may be larger, but generally it is nontrivial to
predict the possible influence of the emitted radiation -- in particular the
coupling to the electromagnetic field complicates the description. For a
discussion of the interaction between electromagnetic fields and
gravitational radiation in an astrophysical context, see for example Refs.\ 
\cite
{Bingham-etal,Brodin-Marklund-Dunsby,Daniel-Tajima,Ignatev,Macdonald-Thorne,Marklund-Brodin-Dunsby,Mendonca-Shukla-Bingham}%
, and references therein.

In the present paper we will study the propagation of gravitational
perturbations in a magnetized plasma, with the direction of propagation
perpendicular to the magnetic field. It turns out that large density
gradients driven by the gravitational perturbation can be generated, even
for small deviations from flat space, provided the cyclotron frequency is
much larger than the plasma frequency. Furthermore, as is well known from
laboratory plasmas (see, e.g., \cite{Wilks89}), moving density gradients can
increase (or decrease) the frequency of electromagnetic wave packets, so
called photon acceleration. The density gradients in our case are
propagating with exactly the speed of light, in contrast to the laboratory
application \cite{Wilks89}. In principle this means that a given photon may
increase its energy by several orders of magnitude, {\em independent of its
initial energy}. Applying our results to gravitational radiation generated
by binary systems, it turns out that the regime of most interest is the
infrared regime. In this case frequency conversion by an order of magnitude
is possible, for a binary system close to merging.

%%%%%%%%%%%%%%%%%%%%%%%%%%%%%%%%%%%%%%%%%%%%%%%%%%%%%%%%% 

\section{Plasma response to a gravitational wave pulse}

%%%%%%%%%%%%%%%%%%%%%%%%%%%%%%%%%%%%%%%%%%%%%%%%%%%%%%%%% 

%%%%%%%%%%%%%%%%%%%%%%%%%%%%%% 

\subsection{Basic equations}

%%%%%%%%%%%%%%%%%%%%%%%%%%%%%% 

The metric of a linearized gravitational wave propagating in the $z$%
-direction can be written as \cite{Landau-Lifshitz} 
\begin{equation}
{\rm d}s^{2}=-{\rm d}t^{2}+[1+h(u)]{\rm d}x^{2}+[1-h(u)]{\rm d}y^{2}+{\rm d}%
z^{2}\ ,
\end{equation}
where we have assumed linear polarization, and $u\equiv z-t$. For an
observer comoving with the time coordinate, the natural frame for
measurements is given by 
\begin{equation}
e_{0}=\partial _{t}\ ,\ e_{1}=\left( 1-{%
%TCIMACRO{\tfrac{1}{2}}%
%BeginExpansion
{\textstyle{1 \over 2}}%
%EndExpansion
}h\right) \partial _{x}\ ,\ e_{2}=\left( 1+{%
%TCIMACRO{\tfrac{1}{2}}%
%BeginExpansion
{\textstyle{1 \over 2}}%
%EndExpansion
}h\right) \partial _{y}\ ,\ e_{3}=\partial _{z}\ .
\end{equation}
It can be shown \cite{Marklund-Brodin-Dunsby} that in such a frame,
Maxwell's equations can be written 
\begin{mathletters}
\begin{eqnarray}
{\boldsymbol{\nabla\cdot}}{\boldsymbol{E}} &=&\rho /\epsilon _{0}\ ,
\label{eq:max1} \\
{\boldsymbol{\nabla\cdot}}{\bf B} &=&0\ ,  \label{eq:max2} \\
\frac{\partial {\boldsymbol{E}}}{\partial t}-{\boldsymbol{\nabla\times}}{%
\boldsymbol{B}} &=&-{\boldsymbol{j}}_{\!_{E}}-\mu _{0}{\boldsymbol{j}}\ ,
\label{eq:max3} \\
\frac{\partial {\boldsymbol{B}}}{\partial t}+{\boldsymbol{\nabla\times}}{%
\boldsymbol{E}} &=&-{\boldsymbol{j}}_{\!_{B}}\ ,  \label{eq:max4}
\end{eqnarray}
where the effective gravitational current densities are defined as 
\end{mathletters}
\begin{mathletters}
\begin{eqnarray}
j_{\!_{E}}^{1} &=&j_{\!_{B}}^{2}={%
%TCIMACRO{\tfrac{1}{2}}%
%BeginExpansion
{\textstyle{1 \over 2}}%
%EndExpansion
}(E^{1}-B^{2})\frac{\partial {h}}{\partial z}\ ,  \label{EffectiveA} \\
j_{\!_{E}}^{2} &=&-j_{\!_{B}}^{1}=-{%
%TCIMACRO{\tfrac{1}{2}}%
%BeginExpansion
{\textstyle{1 \over 2}}%
%EndExpansion
}(E^{2}+B^{1})\frac{\partial {h}}{\partial z}\ ,  \label{EffectiveB}
\end{eqnarray}
and ${\boldsymbol{\nabla}}\equiv (e_{1},e_{2},e_{3})$.

To first order in $h$, the fluid equations become 
\end{mathletters}
\begin{mathletters}
\begin{eqnarray}
\frac{\partial n}{\partial t} + {\boldsymbol{\nabla\cdot}}(n{\boldsymbol{v}}%
) &=& 0 \ , \\
\left( \frac{\partial}{\partial t} + {\boldsymbol{v}}{\boldsymbol{\cdot%
\nabla}} \right)\gamma{\boldsymbol{v}} &=& \frac{q}{m}({\boldsymbol{E}} + {%
\boldsymbol{v}}{\boldsymbol{\times}}{\boldsymbol{B}}) \ ,  \label{eq:mom1}
\end{eqnarray}
where $\gamma \equiv (1 - v_{\parallel}^2)^{-1/2}$, $v_{\parallel} \equiv
v_3 $, and $n = \gamma\tilde{n}$, where $\tilde{n}$ is the proper number
density. These equations holds for each particle species. Note that in
general terms proportional to $v_1h$ and $v_2h$ appear in the equations \cite
{Brodin-Marklund-Dunsby}. Throughout this paper, we will assume that $v_1,
v_2 \ll 1$, and thus neglect terms of order $v_1h$, $v_2h$.

%%%%%%%%%%%%%%%%%%%%%%%%%%%%%%%%%%%%%%%%%%%%%%%%%%%%%%%%%%%%%%%%%%%%% 

\subsection{Electromagnetic fields driven by a gravitational perturbation}

%%%%%%%%%%%%%%%%%%%%%%%%%%%%%%%%%%%%%%%%%%%%%%%%%%%%%%%%%%%%%%%%%%%%% 

From now on we assume $\partial /\partial t\ll \omega _{{\rm c}}$, where $%
\omega _{{\rm c}}\equiv qB/m$ is the cyclotron frequency, for all particle
species (since the gravitational perturbation is assumed to be the driver of
all perturbations this scaling thereby holds for $\partial /\partial t$
acting on all fields). Furthermore, we assume the presence of an external
magnetic field: ${\boldsymbol{B}}_{0}=B_{0}e_{1}$ [where the total field is $%
{\boldsymbol{B}}=(B_{0}+\delta B)e_{1}$]. The electric field takes the form $%
{\boldsymbol{E}}=E_{\perp }e_{2}$.

Looking for solutions driven by the gravitational perturbation, and thus
using $\partial /\partial t=-\partial /\partial z$, we first consider
Faraday's law for $\delta B\ll B_{0}$, which gives 
\end{mathletters}
\begin{equation}
\delta B=-E_{\perp }+hB_{0}  \label{eq:Faraday}
\end{equation}
Next we note that if the excited fields $E_{\perp }$ and $\delta B$ grows
(invalidating $\delta B\ll B_{0}$), the quantity $E_{\perp }+B$ that appears
in the effective current still becomes $E_{\perp }+\delta B=hB_{0}$, and
thereby the above formula holds for arbitrary electromagnetic amplitude.
Taking the time derivative of Ampere's law, using Eq.\ (\ref{eq:Faraday}),
we obtain 
\begin{equation}
\left[ \frac{\partial ^{2}}{\partial t^{2}}-\frac{\partial ^{2}}{\partial
z^{2}}\right] E_{\perp }+\mu _{0}\sum_{i}\frac{\partial j_{\perp (i)}}{%
\partial t}=-2\frac{\partial ^{2}h}{\partial t^{2}}B_{0}\ ,
\label{eq:Ampere}
\end{equation}
where the sum is over particle species, and $j_{\perp }\equiv j_{2}$. For $%
\partial /\partial t=-\partial /\partial z$, the term (explicitly) involving 
$E_{\perp }$ vanishes. The currents are determined by the equation of
motion, noting that the condition $\partial /\partial t\ll \omega _{{\rm c}}$
means that the current contribution from different particle species cancel
to lowest order in an expansion in the operator $\omega _{{\rm c}%
}^{-1}\partial /\partial t$. The equation of motion gives 
\begin{equation}
v_{\parallel }=-\frac{E_{\perp }}{B_{0}+\delta B}  \label{eq:mom2}
\end{equation}
to lowest order. Note that, using Eq.\ (\ref{eq:Faraday}), we can now
approximate the denominator in Eq.\ (\ref{eq:mom2}) by $B_{0}-E_{\perp }$.
The error this approximation introduces will not have any noticeable
effects. This is because $v_{\parallel }$ can only be altered significantly
by the omitted term if $\delta B\approx B_{0}$, but this regime is
inaccessible, since -- from Eq.\ (\ref{eq:mom2}) -- it correspond to
superluminal speeds. From the parallel component of Eq.\ (\ref{eq:mom1}) we
can calculate the first order correction to the induced velocity, which
subsequently determines the current. We obtain 
\begin{equation}
v_{\perp }=-\frac{m}{q}\,\frac{1-v_{\parallel }}{B_{0}-E_{\perp }}\,\frac{%
\partial (\gamma v_{\parallel })}{\partial t}  \label{eq:mom3}
\end{equation}
Furthermore, the continuity equation gives 
\begin{equation}
\delta n=\frac{n_{0}v_{\parallel }}{1-v_{\parallel }}  \label{eq:cont}
\end{equation}
where we have divided the density into a perturbed and an unperturbed part, $%
n=n_{0}+\delta n$.

>From (\ref{eq:Ampere}) and the relations above we can thus determine the
induced velocity and density in terms of the metric perturbation $h$. The
result (for all particle species) is 
\begin{mathletters}
\label{eq:perturbations}
\begin{eqnarray}
v_{\parallel } &=&\frac{1-\left( 1-{\cal H}\right) ^{2}}{1+\left( 1-{\cal H}%
\right) ^{2}}\ ,  \label{eq:parallelvelocity} \\
\delta n &=&\frac{n_{0}}{2}\left[ \frac{1}{\left( 1-{\cal H}\right) ^{2}}-1%
\right] \ ,  \label{eq:deltadensity}
\end{eqnarray}
where ${\cal H}\equiv 2h/\sum_{i}(\omega _{{\rm p}(i)}^{2}/\omega _{{\rm c}%
(i)}^{2})$, and $\omega _{{\rm p}(i)}\equiv (q_{(i)}^{2}n_{0}/\epsilon
_{0}m_{(i)})^{1/2}$ is the plasma frequency for the unperturbed plasma
species $i$. Thus it is clear that even a moderate or small value of the
gravitational perturbation may cause significant density perturbation,
provided the plasma is strongly magnetized in the sense that $%
\sum_{i}(\omega _{{\rm p}(i)}^{2}/\omega _{{\rm c}(i)}^{2})\ll 1$. This is
because the fast magnetosonic (or compressional Alfv\'{e}n) wave fulfills
approximately the same dispersion relation as the gravitational wave, with
the mismatch being proportional to $\sum_{i}(\omega _{{\rm p}(i)}^{2}/\omega
_{{\rm c}(i)}^{2})$ \cite{Regime}. The divergence that occur for ${\cal H}%
\rightarrow 1$ is clearly unphysical, and it's removal will be discussed in
the next subsection.

For future considerations it will also be useful to have the relation
between the relative magnetic field perturbation and the relative density
perturbation. When $\left|\delta B\right| \gg \left| hB_{0}\right| $, which
is the case of most interest, the last term of Eq.\ (\ref{eq:Faraday}) can
be neglected and the desired relation can be derived by combining the
resulting formula with Eqs.\ (\ref{eq:mom2}) and (\ref{eq:cont}). The simple
result is 
\end{mathletters}
\begin{equation}
\frac{\delta n}{n_0} = \frac{\delta B}{B_0}  \label{eq:pertfrac}
\end{equation}

%%%%%%%%%%%%%%%%%%%%%%%%%%%%%%%%%%%%%%% 

\subsection{Removal of the divergence}

%%%%%%%%%%%%%%%%%%%%%%%%%%%%%%%%%%%%%%%
\label{sec:divergence}

The purpose in this subsection is to explain the reason for the occurrence
of divergence when ${\cal H}$ approaches unity, and to discuss various
modifications of the assumptions that lead to a more physical behavior. From
Eq.\ (\ref{eq:mom3}) we note that for infinitesimal velocity perturbations, $%
v_{\perp }$ (and thereby $j_{\perp }$) depends linearly on $v_{\parallel }$,
but for large parallel velocities, in particular when $v_{\parallel
}\rightarrow 1$, $v_{\perp }$ remains finite due to the factor $%
1-v_{\parallel }$. From Eq.\ (\ref{eq:Ampere}) it is thus clear that we {\em %
cannot} have a stationary solution where $E_{\perp }$ depend only on $z-t$
for large enough $h$, and from Eq.\ (\ref{eq:parallelvelocity}) we see that
this limit for the gravitational perturbation is reached when ${\cal H}$
becomes unity. Basically, the physical reason is the following: In vacuum
the electromagnetic and gravitational modes obey the same dispersion
relation, and therefore -- due to the mode coupling provided by the
unperturbed magnetic field -- the system evolves in a non-stationary way. In
particular gravitational wave energy may be continuously converted into
electromagnetic wave energy, as will be examined in more detail below. In
the presence of a plasma, however, the induced currents change the
dispersion relation of the electromagnetic wave, and the resulting detuning
of the modes saturate the conversion of energy between them, making a steady
state solution (in a frame moving with the velocity of light) possible in
principle. For a strongly magnetized plasma, on the other hand, the induced
plasma currents cannot grow continuously with $h$, as we have seen above.
For sufficiently high gravitational amplitude this means that the plasma
currents are of little significance, practically the plasma appears as
vacuum for ${\cal H}\geq 1$, and in particular solutions depending only on $%
z-t$ are impossible. This conclusion is {\em not} dependent on the absence
of thermal effects in our calculations in section II\ B. Generally the
addition of thermal motion only modifies our expressions (\ref
{eq:perturbations}) by a factor of the order $1+(v_{t}/c)^{2}$, where $v_{t}$
is the thermal velocity. In particular, the divergence of (\ref
{eq:deltadensity}) still occurs for a finite value of ${\cal H}$.

On the other hand, it is clear that our omission of the back reaction of the
electromagnetic wave on the gravitational pulse in principle could change
this picture, since obviously certain components of the energy momentum
tensor also diverges when ${\cal H}\rightarrow 1$, implying that the
gravitational wave amplitude could indeed be diminished due to the influence
of the generated EM-wave. The effects of the selfconsistent gravitational
field caused by the plasma perturbations are discussed in the Appendix, but
will be omitted here since it turns out that the backreaction on the
gravitational wave is negligible in the application to be discussed in this
article.

Since it is clear that for ${\cal H}\geq 1$ the generated currents cannot
stop the growth of the EM-wave, we simplify the picture from now on by
putting the density to zero and thus totally ignoring the plasma effects.
The general solution to Eq.\ (\ref{eq:Ampere}) for the electric field in the
presence of a monochromatic gravitational wave $h=\widetilde{h}\cos [k(z-t)]$
can then be written 
\begin{eqnarray}
\delta B &=&E_{\perp }=%
%TCIMACRO{\tfrac{1}{2}}%
%BeginExpansion
{\textstyle{1 \over 2}}%
%EndExpansion
k(C_{z}z+C_{t}t)B_{0}\widetilde{h}\sin [k(z-t)]  \nonumber \\
&&+E_{+}(z-t)+E_{-}(z-t)\ ,  \label{eq:magneticpert}
\end{eqnarray}
where $C_{z}+C_{t}=1$ and $E_{+}$ and $E_{-}$ are arbitrary functions. For
an initial value problem where the plasma is unperturbed in the absence of
the pulse $C_{z}=0,$ $C_{t}=1$ and $E_{+}=E_{-}=0$, i.e. the electromagnetic
amplitude grows linearly with time. For a boundary value problem, on the
other hand, where the external magnetic field $B_{0}$ occupies a region $%
z\geq 0$ and there is a gravitational wave but no EM-waves propagating into
\ the magnetized region, clearly $C_{z}=1,$ $C_{t}=0$ and $E_{+}=E_{-}=0$,
i.e. we have a linear spatial growth instead. For the applications to be
discussed later on we will be interested in a situation where $B_{0}$ is not
necessarily static. We thus note that qualitatively the solution given by
Eq.\ (\ref{eq:magneticpert}) still applies for a quasi-static situation,
i.e. where the dependence of $B_{0}$ on time is slow enough such that the
electric fields $E$ associated with the time variations fulfills $E/B_{0}\ll
1$.

In principle we can achieve very large EM-wave amplitudes also when we
abandon the specific solutions depending on $z-t$. However, since the growth
is only linear in $t$ and/or $z$, apparently we need large times/distances
of coherent interaction. For a boundary value problem we can roughly define
the effective distance of interaction as $z_{{\rm eff}}$ 
\begin{equation}
\delta B_{{\rm max}}\simeq z_{{\rm eff}}B_{0,{\rm char}}h_{{\rm char}%
}^{\prime }  \label{eq:maxpert}
\end{equation}
where the index ${\rm char}$ denotes the characteristic values of the
various quantities in the region of interest, and the prime denote
differentiation with respect to the argument.

To summarize: Eq. (\ref{eq:deltadensity}) have a class of physically sound
solutions, but also unphysical ones with the property $\delta n\rightarrow
\infty $ as ${\cal H}\rightarrow 1.$ The singular behavior is caused by the
insistence to look for solutions that moves with a specific velocity,
together with the omission of the selfconsistent gravitational field from
the plasma perturbations. The divergent solutions can be removed either by
considering a boundary or an initial value problem, as discussed in this
subsection, or by considering the backreaction of the plasma perturbations
on the gravitational wave, as discussed in the Appendix. The alternative
considered here is the most relevant one with regard to astrophysical
applications. Real astrophysical systems have finite distances of
interaction between gravitational waves and plasma waves, that can be
estimated on physical grounds. Thus when estimating the maximum magnetic
field perturbation that can be produced by a gravitational wave in a given
situation, we can in principle apply solutions (\ref{eq:deltadensity})
together with (\ref{eq:pertfrac}) but we must note the upper bound for $%
\delta B_{{\rm max}}$ that exists for a given $z_{{\rm eff}}$ and is given
by Eq. (\ref{eq:maxpert}).

%%%%%%%%%%%%%%%%%%%%%%%%%%%%%%%%% 

\section{Photon frequency shift}

%%%%%%%%%%%%%%%%%%%%%%%%%%%%%%%%% 
\label{sec:photonacc}

We now consider the effect of the gravitational wave perturbations on high
frequency photons in a plasma. For simplicity we assume that the photons
propagate parallel to the gravitational waves and let them be represented by
the vector potential ${\boldsymbol{A}}={\boldsymbol{\tilde{A}}}\exp ({\rm i}%
\,\theta )+{\rm c.c.}$, where ${\rm c.c.}$\ stands for complex conjugate.
Making the approach of geometrical optics \cite{Landau-Lifshitz}, the wave
number $k\equiv \partial _{z}\theta $ and frequency $\omega \equiv -\partial
_{t}\theta $ satisfies some local dispersion relation $\omega =W(z,t,k)$.
The amplitude of the vector potential is assumed small and by high frequency
photons we mean $\omega \gg \omega _{{\rm p}(i)},\omega _{{\rm c}(i)}$.

Due to the gravitational waves the plasma has a background of possibly large
fields $\delta n$, $v_{\parallel }$, $\delta B$, and $E_{\perp}$ all being
functions of $z - t$ and varying on a time and length scale much longer than
that of ${\boldsymbol{A}}$.

Since $\omega \gg \omega _{{\rm c}(i)}$ the high frequency pulse
approximately behaves as if the plasma is unmagnetized. The equation of
motion linearized in the high frequency (hf) variables reads 
\begin{equation}
\left[ \frac{\partial}{\partial t} + v_{\parallel}\frac{\partial}{\partial z}
\right] {\boldsymbol{v}}_{(i)}^{{\rm hf}} = \frac{-q_{(i)}}{\gamma m_{(i)}}%
\left( \frac{\partial{\boldsymbol{A}}}{\partial t} + v_{\parallel}\frac{%
\partial{\boldsymbol{A}}}{\partial z}\right) \ ,
\end{equation}
and thus ${\boldsymbol{v}}_{(i)}^{{\rm hf}} = -q_{(i)}{\boldsymbol{A}}%
/\gamma m_{(i)}$, where the large scale variations have been neglected. The
induced high frequency current is therefore ${\boldsymbol{j}}^{{\rm hf}} =
-\omega_{{\rm p}}^2{\boldsymbol{A}}/\mu_0$, where the plasma frequency is $%
\omega_{{\rm p}} \equiv (\sum_{i}q_{(i)}^{2}n/\epsilon_0m_{(i)}\gamma)^{1/2}$%
. Taking the time derivative of Ampere's law gives the following wave
equation for the photons 
\begin{equation}
\left[ \frac{\partial^2}{\partial t^2} - \frac{\partial^2}{\partial z^2} +
\omega_{{\rm p}}^2 \right]{\boldsymbol{A}} = 0 \ .
\end{equation}
We recognize the dispersion relation as $\omega = [ k^2 + \omega_{{\rm p}%
}^2(z - t) ]^{1/2}$, where we assume that the variations in the plasma
frequency are determined from Eqs.\ (\ref{eq:perturbations}) together with $%
\gamma = (1 - v^2_{\parallel})^{-1/2}$.

The change in the wave number and frequency as the wave propagates through
the nonuniform and time varying media with velocity $v_{g}=\partial\omega
/\partial k$ is given by the ray equations 
\begin{equation}
\frac{dk}{dt} = -\frac{\partial W}{\partial z} \ , \quad \frac{d\omega}{dt}
= \frac{\partial W}{\partial t} \ .
\end{equation}
Note that $W$ is a function of $z - t$ and introduce coordinates, $\xi = z -
v_gt$, $\tau = t$ locally moving with the photons, i.e.\ it should be
understood that $v_{g} = v_{g}(\tau = \tau_0)$ for some $\tau_0$. Then, in a
small neighborhood of $\tau_0$ it holds that $d\omega/d\tau = -\partial
W/\partial\xi$. Using $\partial_{\xi} = (1 -v_g)^{-1} \partial_{\tau}$, this
can be integrated from time 1 to 2 (which need not be a small interval),
noting that $1 - v_g \approx \omega_{{\rm p}}^2/2\omega^2$. The result is 
\begin{equation}
\frac{\omega_1}{\omega_2} = \frac{{\omega_{{\rm p 1}}^2}}{{\omega_{{\rm p 2}%
}^2}} \ ,  \label{eq:freqshift}
\end{equation}
where the indices $1$ and $2$ denote the values at $\tau_1$ and $\tau_2$,
respectively. An interesting aspect of Eq.\ (\ref{eq:freqshift}) is that the
frequency conversion factor $N = \omega_1/\omega_2$ is independent of the
frequency regime of the EM-wave. Thus, in principle, x-rays can be turned
into gamma rays, just as well as infra-red waves can be converted into the
visible regime. This is in contrast to laser excited wake fields \cite
{Wilks89}, where efficient frequency shifts can only take place provided the
frequency of the converted pulse roughly lies in the same frequency regime
as the exciting laser pulse. The reason for the difference is that the
density gradients propagates with exactly the speed of light in our case,
whereas, naturally, the corresponding velocity is slightly less than $c$ in
the laboratory experiments. The necessary distance of acceleration for a
given conversion factor $N$ is proportional to $\omega^2$, however, and this
puts certain limits for the applicability to the highest frequency regimes.

%%%%%%%%%%%%%%%%%%%%%%%%%% 

\section{Example}

%%%%%%%%%%%%%%%%%%%%%%%%%%
\label{sec:example}

We have found that large density perturbations traveling with the velocity
of light can be induced by small gravitational wave perturbations, provided
the cyclotron frequency is much larger than the plasma frequency, as
described by Eqs.\ (\ref{eq:perturbations}). Furthermore, photons
propagating in a moving density gradient can undergo frequency up-conversion
(or down-conversion), as described by Eq.\ (\ref{eq:freqshift}). In
principle the effects can be large, even for a moderate deviation from flat
space-time. It is not yet clear that the predicted frequency conversion can
be observed during reasonable conditions, however, and our aim in this
section is to provide estimates to shed light on this question. In this
section we reinstate the speed of light in all expressions.

As a source of gravitational radiation we consider a binary system. At least
one of the objects should have a moderate to strong magnetic field (in order
to make the parameter $\sum_{i}(\omega _{{\rm p}(i)}^{2}/\omega _{{\rm c}%
(i)}^{2})$ small), and the objects should be compact (as to make the
gravitational wave frequency and amplitude before merging large). Thus, for
definiteness (and calculational simplicity due to symmetries) we assume that
the system consists of two neutron stars of equal mass $M_{\odot }$,
separated by a distance of $40\ R_{S}$, where $R_{S}=2GM_{\odot}/c^2 \approx
3$ km. Furthermore, the surface magnetic field of each neutron star is
assumed to be $10^{6}$ T. For the unperturbed plasma density profile, see
Fig.\ \ref{fig:binary}.

The area surrounding the binary system can loosely be divided into three
regions (Fig.\ \ref{fig:binary}). The interval $20\ R_{S}-30\ R_{S}$ from
the center of mass (CM) roughly constitutes region I, which is the region
where most of the gravitational energy is gained by the EM-wave. Using a
Newtonian approximation, with $d=\alpha R_{S}$ and $r=\beta R_{S}$, it is
straightforward to show that $|h|\sim (2\alpha \beta )^{-1}$, where $d$ is
the separation distance between the binary objects and $r$ is the
observation distance from the center of mass of the system. In order to
obtain an estimate of the amplitude of the generated EM-wave, we combine the
above expression for the gravitational wave amplitude with Eq.\ (\ref
{eq:maxpert}) and the data given above. The result is 
\begin{equation}
\frac{\delta B}{B_{0}}\sim 7\times 10^{-5}
\end{equation}
at the of end region I. In region II (approximate interval $30\ R_{S}-3500\
R_{S}$ from the CM) $\delta B/B_{0}$ is still small, and -- as seen by Eq.\ (%
\ref{eq:pertfrac}) -- the relative density perturbation is thereby small as
well, which limits the frequency conversion effect in this region. However,
the gravitationally induced EM-wave suffers spherical attenuation, whereas
the unperturbed magnetic field is that of a dipole, and consequently the
relative density perturbation grows quadratically with distance. The end of
region II is defined as the necessary distance to make $\delta B/B_{0}$ \ of
the order unity due to this increase. (For pulsars with period longer than $%
35$ ms, region I and II lies in the near zone, and thus the unperturbed
magnetic field indeed decays cubically in the region of interest, although
the unperturbed field becomes a radiation field outside the light cylinder
of the pulsar.) In region III (approximate interval $3500\ R_{S}-10^{6}\
R_{S}$), the relative density perturbation is appreciable, and thus the main
frequency conversion occurs here \cite{RegionIII}. 
%Assuming that the EM-wave excitation starts at a distance ...  
%from the center of mass, the effective unperturbed magnetic field is roughly  
%..., and 

At the beginning of region III the relative density perturbation is $\delta
n/n_{0}\sim 1$, in agreement with Eq.\ (\ref{eq:pertfrac}). An EM-wave with
initial frequency $\omega \equiv \omega _{{\rm min}}=10^{12}\ {\rm rad}/{\rm %
s}$ can move from a density minimum to a density maximum during a
``laboratory system distance'' $L_{{\rm freq}}=cT_{{\rm freq}}=cL_{{\rm grad}%
}/(c-v_{g})\sim \omega _{\max }^{2}L_{{\rm grad}}/\omega _{{\rm p}}^{2}$,
where $L_{{\rm grad}}$ is a typical density gradient scale length. For
definiteness we assume that the pulsars have periods of the order of 350 ms,
in which case $\delta B/B_{0}$ may increase to $\delta B/B_{0}\sim 10$ for
the most of region III. In our example the maximum frequency magnification $%
N $ thus is 
\begin{equation}
N=\frac{\omega _{{\rm max}}}{\omega _{{\rm min}}}=\frac{\omega _{{\rm p},%
{\rm max}}^{2}}{\omega _{{\rm p},{\rm min}}^{2}}\sim 10\ .
\label{eq:magnification}
\end{equation}
Inserting $\omega _{{\rm max}}=10^{13}\ {\rm rad}/{\rm s}$, and letting $%
\omega _{{\rm p},{\rm max}}^{2}=10^{11}\ {\rm rad}/{\rm s}$ (corresponding
to $n_{0}\simeq 10^{12}\ {\rm cm}^{-3}$) we obtain $L_{{\rm freq}}\simeq
10^{6}R_{{\rm S}}$, $\ $i.e. the acceleration can take place within region
III. Strictly applying our one-dimensional calculations of Sec.\ \ref
{sec:photonacc} means that frequency up-converted EM-waves will be
down-converted and vice versa, if the gravitational source and the induced
density perturbation are indeed periodic. In our example, on the other hand,
the successive frequency conversion effects will decrease with the distance
from the source, and thus for an earth based observer the radiation
generated in region III should show periodic up- and down-conversions. The
frequency conversion ratio of Eq.\ (\ref{eq:magnification}) is of course a
maximum value of our example, that occurs for radiation generated at a
density extremum, but all radiation generated in region III will be up- or
down converted with a factor in the interval $1-N$, and consequently the
effect should be observable provided the object is close enough for
radiation generated in region III, in the approximate frequency interval $%
10^{11}\ {\rm rad}/{\rm s}\leq \omega \leq 10^{14}\ {\rm rad}/{\rm s}$, to
be detected, where the upper limit is imposed by the fact that the system
has a finite distance of interaction. If we try to increase the interaction
efficiency by considering higher plasma densities the electromagnetic wave
damping due to Thomson scattering becomes prohibitive\cite{RegionIII}.

%%%%%%%%%%%%%%%%%%%%%%%%%%%%%%%%%%% 

\section{Summary and discussion}

%%%%%%%%%%%%%%%%%%%%%%%%%%%%%%%%%%% 

We have considered the generation of traveling density perturbations in a
magnetized plasma induced by gravitational radiation. Provided $%
\sum_{i}(\omega _{{\rm p}(i)}^{2}/\omega _{{\rm c}(i)}^{2}) \ll 1$,
significant density perturbations, i.e. $\delta n/n_{0}\sim 1$, can be
induced even by a small gravitational wave with $h\ll 1$, provided ${\cal H}
\sim 1$. Basically the large effect is possible because of the approximate
agreement of the dispersion relations between the fast magnetosonic and
gravitational modes in the regime $\sum_{i}(\omega _{{\rm p}(i)}^{2}/\omega
_{{\rm c}(i)}^{2}) \ll 1$, which in turn allows for a long distance of
coherent interaction.

In order to find a mechanism where the induced density perturbations may
give rise to earth-based observational effects, we have studied frequency
conversion of electromagnetic wave packets traveling in the moving density
gradients. The formula (\ref{eq:freqshift}), relating the frequency of the
wave packet for two different positions in the moving density profile, is in
conceptual agreement with the corresponding results of Ref.\ \cite{Wilks89},
which considered an analogous situation but where the density perturbation
was due to plasma oscillations traveling with a phase velocity slightly less
than the speed of light $c$. In our case the gradients move with exactly $c$%
, however, and thereby the maximum frequency conversion factor $N$ does not
decrease with the initial frequency [as for conventional photon
acceleration], {\em in principle} allowing for up-conversion even of $\gamma 
$-rays.

The idealizations made in Secs.\ II and \ref{sec:photonacc} is a somewhat
too strong for our results to be directly applicable to a situation of
astrophysical relevance. In particular, we cannot consider the unperturbed
plasma as homogeneous and the geometry as one-dimensional when making
estimates. In our example with a binary system as a source of gravitational
radiation, we have thus been forced to divide the neighborhood of the system
into three regions: Region I where most of the energy transfer into
electromagnetic wave energy occurs, region II where the relative density
perturbation grows, and region III where the frequency conversion takes
place. In order to describe the physics in region I adequately we must
abandon solutions that depend on $z-ct$ only, and the basis for this has
been discussed in Sec.\ \ref{sec:divergence}. By making estimates based on
our analytical calculations, we conclude that the gravitational waves
emitted by a system of binary pulsars close to merging may result in
periodic frequency up- and down-conversions of electromagnetic radiation in
the infrared part of the spectrum. The frequency of the up- and
down-conversions coincides with the gravitational wave frequency, i.e.\ it
is twice the orbital frequency.

\section{Appendix}

In this appendix we are going to investigate the regime of validity for the
multi-component test fluid approach. Normally we think that by continually
decreasing the parameters proportional to the unperturbed energy density, at
some point the fluid in an external gravitational field can be treated as a
test fluid. In our case the situation is not quite that simple, since we can
decrease the electromagnetic ($\propto B_{0}^{2}$) and the rest mass energy
density ($\propto n_{0}$)at the same rate keeping $\sum_{i}(\omega _{{\rm p}%
(i)}^{2}/\omega _{{\rm c}(i)}^{2})$ constant. Since our solution in section
II B has a diverging energy-momentum tensor whenever ${\cal H}\equiv
2h/\sum_{i}(\omega _{{\rm p}(i)}^{2}/\omega _{{\rm c}(i)}^{2})\rightarrow 1$%
, clearly we cannot justify the test fluid approach simply by assuming a
sufficiently low unperturbed energy density. To shed light on the physical
effects due to the selfconsistent gravitational field, we will first
consider the linearized theory. This will provide a guide for making
estimates of the regime of validity of our (nonlinear) test matter solution
in section II B, and also makes it possible to justify the omission of
selfconsistent gravitational effects in section IV .

We divide all quantities into an unperturbed part (i.e. the value in the
absence of the gravitational perturbation) and a perturbed part. We note
that the only variables that are nonzero in the unperturbed state are the
density ($=n_{0}$), the magnetic field ($=B_{0}e_{1}$) and the metric ($%
=\eta _{\mu \nu }$). It should be emphasized that in addition to the direct
effect on the dispersion relation from the matter, which we will study
below, there is also an indirect contribution (that will be omitted here) to
the dispersion relation from the background curvature produced by the
(unperturbed) matter. In the regime where the gravitational wave length is
much shorter than the background curvature, however, the shortwave
approximation can be applied, which imply that these two effects can be
studied separately and their contribution to the dispersion relation of the
gravitational wave can be added, see e.g. Ref. \cite{Grischuk80}. In the
above scenario (provided thermal effects are still neglected) the only
effects from the gravitational wave on the plasma perturbations are from the
effective currents in (\ref{EffectiveA})-(\ref{EffectiveB}), where, in the
present case, we have $j_{\!_{E}}^{2}=-j_{\!_{B}}^{1}=-(1/2)B_{0}(\partial {%
h/}\partial z{)}$ and the other components are zero. Thus using Maxwells
eqs. and the the set of fluid equations for each particle species and the
same approximations as in section II (but avoiding the ansatz $\partial
/\partial t=\partial /\partial z$) we will obtain a wave equation for the
fast magnetosonic wave, modified from the standard textbook form by allowing
for an arbitrary value of $\sum_{i}(\omega _{{\rm p}(i)}^{2}/\omega _{{\rm c}%
(i)}^{2})$, and with a gravitational ''source term'' due to the effective
gravitational currents above. The result is 
\begin{equation}
\left( \frac{\partial ^{2}}{\partial t^{2}}+\frac{C_{A}^{2}}{1+C_{A}^{2}}%
\frac{\partial ^{2}}{\partial z^{2}}\right) \delta B=2\frac{\partial ^{2}h}{%
\partial t^{2}}B_{0}  \label{Mslin}
\end{equation}
where we have introduced the Alfv\'{e}n velocity $C_{A}=\left(
\sum_{i}(\omega _{{\rm p}(i)}^{2}/\omega _{{\rm c}(i)}^{2})\right) ^{1/2}$%
.(Note that $C_{A}$ may be larger than unity, but, as can be seen above, the
actual magnetosonic wave velocity is smaller or equal to $C_{A}.$) The
system is closed selfconsistently by Einstein's field equations, which,
after linearization reduces to (cf Eq. 4.9 in Ref. \cite{Grischuk80}) 
\begin{equation}
\left( \frac{\partial ^{2}}{\partial t^{2}}+\frac{\partial ^{2}}{\partial
z^{2}}\right) h=16\pi G[T_{11}-T_{22}]_{{\rm lin}}=\frac{16\pi G}{\mu _{0}}%
B_{0}\delta B  \label{GWlin}
\end{equation}
where ${\rm lin}$ stands for ''linear part of''. It is simple to combine
Eqs. (\ref{Mslin}) and (\ref{GWlin}) into a single wave equation for the
coupled fast magnetosonic and gravitational mode. However, it is probably
more illustrative to proceed by considering the corresponding dispersion
relation. Making a plane wave ansatz, $\delta B=\widetilde{\delta B}\exp
[i(kz-\omega t)]$ and $h=\widetilde{h}\exp [i(kz-\omega t)]$, we directly
find the dispersion relation: 
\begin{equation}
\omega ^{2}-k^{2}=\frac{32\pi GB_{0}^{2}}{\mu _{0}}\left( \frac{\omega ^{2}}{%
\omega ^{2}-k^{2}C_{A}^{2}/(1+C_{A}^{2})}\right)   \label{GWdisp-rel}
\end{equation}
from Eqs. (\ref{Mslin}) and (\ref{GWlin}). Thus the presence of matter
causes a phase velocity $\omega /k>1$ and a group velocity $d\omega /dk<1$.
A further consequence is that the gravitational wave also becomes
dispersive. Apparently the relation between $\widetilde{\delta B}$ and $%
\widetilde{h}$ is 
\begin{equation}
\widetilde{\delta B}=B_{0}\widetilde{h}\left( \frac{\omega ^{2}}{\omega
^{2}-k^{2}C_{A}^{2}/(1+C_{A}^{2})}\right)   \label{dB-fnk-h}
\end{equation}
where the omission of the selfconsistent gravitational field is a valid
approximation only if we can use the vacuum dispersion relation $\omega
^{2}-k^{2}=0$ as an approximation instead of Eq. (\ref{GWdisp-rel}) when
calculating $\widetilde{\delta B}$ from (\ref{dB-fnk-h}). From now on we
will focus on the regime $C_{A}\gg 1$, which make the magnetosonic phase
velocity close to unity. Since the (typically small) right hand side of (\ref
{GWdisp-rel}) now must be compared to the small phase velocity difference of
the (uncoupled) magnetosonic and gravitational waves, the condition for
omitting the selfconsistent gravitational field is significantly stronger if
one should get an approximately correct {\it magnetic field}, and not just a
small contribution from the right hand side in the dispersion relation (\ref
{GWdisp-rel}). For $C_{A}\gg 1$ the condition for omitting the
selfconsistent gravitational field, and still obtaining an approximate
expression for $\widetilde{\delta B},$ becomes: 
\begin{equation}
\frac{32\pi GB_{0}^{2}}{\mu _{0}}\ll \frac{\omega ^{2}}{C_{A}^{4}}
\label{lin-cond}
\end{equation}
The above validity condition is obtained by comparing the magnetic field
obtained from the full selfconsistent dispersion relation and its vacuum
approximation. A much simpler way to arrive at the same condition as in (\ref
{lin-cond}) is to demand that the relative contribution from the energy
momentum tensor terms in Einstein's equations should be much smaller than
the relative velocity difference between the magnetosonic and gravitational
waves. The advantage with this latter formulation of the validity condition
is that it can be easily applied also when the relation between $\delta B$
and $h$ as well as the expression for the energy momentum tensor are
nonlinear. Adopting this condition for omitting the selfconsistent
gravitational field when the plasma response to the metric perturbation is
nonlinear we write 
\begin{equation}
32\pi G\max (\delta T)\ll \frac{\omega _{{\rm char}}^{2}}{C_{A}^{2}}h_{{\rm %
char}}  \label{nl-cond}
\end{equation}
where $\max (\delta T)$ denotes the maximum deviation from the unperturbed
value of the perturbed energy momentum tensor for any of its components, and
the index ''char'' denotes the characteristic value of the gravitational
wave frequency and metric perturbation, respectively. For the regime when
Eq. (\ref{nl-cond}) is violated, obviously our solution in section II B must
be modified to take the selfconsistent gravitational field into account, and
this may result in new types of solutions describing, for example, nonlinear
solitary gravitational pulses. This problem is outside the scope of our
present article, however. We note that our example in section IV, fulfills
the validity condition (\ref{nl-cond}) with a margin of several orders of
magnitude.

%%%%%%%%%%%%%%%%%%%%%%%%%%%%%%

\newpage

%%%%%%%%%%%%%%%%%%%%%%%%%%%%%%%%%%%%%%%%%%
\begin{figure}[tbp]
\caption{The neighborhood of the binary system is divided into three
regions: Region I ($20R_S - 30R_S$), region II ($30R_S - 3500R_S$), and
region III ($3500R_S - 10^6R_S$). In regions I and II we are situated in the
near zone of the magnetic field of the pulsar. Thus the plasma density is
low, and we assume that the plasma particles do not interfere with the
approximations made in the example. For this to be true, ${\cal H} \gg 1$
must hold in region I, which is satisfied even for very high densities.
Furthermore $\sum_{i}(\protect\omega _{{\rm p}(i)}^{2}/\protect\omega_{{\rm c%
}(i)}^{2}) \geq 1$ should apply in region II, which is fulfilled for $n_0
\leq 10^6 \ {\rm cm}^{-3}$. In region III, which is mainly outside the light
cylinder of the pulsar, we assume the plasma density $n_0$ to be of the
order of $10^{12} \ {\rm cm}^{-3}$.}
\label{fig:binary}
\end{figure}
%%%%%%%%%%%%%%%%%%%%%%%%%%%%%%%%%%%%%%%%%%


\begin{references}
\bibitem{Regime}  Note that in the regime of interest, the linear dispersion
relation of the fast magnetosonic wave is $\omega^2 = k^2c^2/[1 +
\sum_i(\omega_{{\rm p}(i)}^2/\omega_{{\rm c}(i)}^2)]$.
\bibitem{RegionIII}  For the assumed electron number density $n_{0}=10^{12}/%
{\rm cm}^{3}$, the mean free path due to Thomson scattering is roughly $%
5\times 10^{6}$ $R_{{\rm S}}$. Thus, the outer boundary imposed by Thomson
scattering is larger than the size of region III.
\end{references}
\end{document}